\documentclass[%
 aps,
pra,%
 amsmath,amssymb,
 reprint,%
]{revtex4-2}

\usepackage{graphicx}
\usepackage{dcolumn}
\usepackage{bm}
\usepackage{hyperref}

\begin{document}

\title{Directional dark-field implicit x-ray speckle tracking using an anisotropic-diffusion Fokker--Planck equation}

\author{Konstantin M. Pavlov}
\email{konstantin.pavlov@canterbury.ac.nz}
\affiliation{School of Physical and Chemical Sciences, University of Canterbury, Christchurch, New Zealand}
\affiliation{School of Physics and Astronomy, Monash University, Victoria 3800, Australia}
\affiliation{School of Science and Technology, University of New England, NSW 2351, Australia}

\author{David M. Paganin, Kaye S. Morgan}
\affiliation{School of Physics and Astronomy, Monash University, Victoria 3800, Australia}

\author{Heyang (Thomas) Li}
\affiliation{School of Mathematics and Statistics, University of Canterbury, Christchurch, New Zealand}

\author{Sebastien Berujon}
\affiliation{Instituto COPPEAD de Administra{\c{c}}\~ao, Universidade Federal do Rio de Janeiro, Rio de Janeiro, Brazil}

\author{Laur\`{e}ne Qu\'{e}not, Emmanuel Brun}
\affiliation{Inserm UA7 STROBE, Universit\'{e} Grenoble Alpes, 38000 Grenoble, France}

\date{\today}

\begin{abstract}
When a macroscopic-sized non-crystalline sample is illuminated using coherent x-ray radiation, a bifurcation of photon energy flow may occur.  The coarse-grained complex refractive index of the sample may be considered to attenuate and refract the incident coherent beam, leading to a coherent component of the transmitted beam.  Spatially-unresolved sample microstructure, associated with the fine-grained components of the complex refractive index, introduces a diffuse component to the transmitted beam.  This diffuse photon-scattering channel may be viewed in terms of position-dependent fans of ultra-small-angle x-ray scatter.  These position-dependent fans, at the exit surface of the object, may under certain circumstances be approximated as having a locally-elliptical shape.  By using an anisotropic-diffusion Fokker--Planck approach to model this bifurcated x-ray energy flow, we show how all three components (attenuation, refraction and locally-elliptical diffuse scatter) may be recovered.  This is done via x-ray speckle tracking, in which the sample is illuminated with spatially-random x-ray fields generated by coherent illumination of a spatially-random membrane.  The theory is developed, and then successfully applied to experimental x-ray data.    
\end{abstract}
\maketitle

\section{Introduction}

Speckle-based phenomena and associated measurement techniques, for both radiation and matter wave fields, appear in numerous optical-physics settings.  Examples include speckle arising from the passage of coherent light through spatially random screens \cite{hariharan1971}, speckle interferometry \cite{labeyrie1970}, photon correlation spectroscopy \cite{pecora1985}, speckle correlography \cite{fienup1988}, Ronchigram aberrometry \cite{lupini2010}, fluctuation microscopy \cite{gibson2007}, ghost imaging \cite{padgett2017introduction}, vortex networks associated with fully developed speckle \cite{OHolleran2008}, turbulence in time-dependent optical speckles \cite{alperin2019},  speckled microdiffraction patterns arising from focused electron beams that scatter from amorphous materials \cite{rodenburg1988} and speckled cross-spectral densities arising from modern undulator sources \cite{paganinsanchezdelrio2019}.  This list is far from complete.  For a broad overview, see e.g.~the book by Goodman \cite{GoodmanSpeckleBook}. 

Our specific focus is x-ray speckle \cite{sutton1991}.  Again, the breadth of x-ray speckle phenomena is extensive.  Techniques employing x-ray speckle include x-ray photon correlation spectroscopy \cite{shpyrko2007}, x-ray ghost imaging \cite{Yu2016fourier,pelliccia2016experimental}, x-ray near-field speckle analysis \cite{cerbino2008}, x-ray particle image velocimetry \cite{LeeKim2003}, analysis of the x-ray scattering of focused probes from disordered materials \cite{clark1983}, x-ray coherent diffractive imaging \cite{miao1999} and analysis of x-ray speckles arising from coherent surface diffraction \cite{pierce2009}.  One particular use of x-ray speckles is the core topic of the present paper: x-ray speckle tracking \cite{berujon2012,morgan2012}. 

Recent years have seen the emergence of new x-ray imaging approaches that tap into x-ray phase information to reveal weakly-absorbing samples, and access an x-ray `dark-field' signal that reveals the location of spatially-unresolved microstructure within the sample. The first phase and dark-field imaging approach used either crystals or multiple gratings to `analyze' the x-ray wavefield downstream of the sample \cite{endrizzi2018}. More recently, high-resolution approaches without an analyzer grating have been shown, instead of using a high-resolution detector to directly capture how the introduction of the sample alters the periodic illumination produced by the primary grating.  An x-ray phase shift introduced by the sample upon the x-ray wavefield will locally transversely shift or warp the image of the grating pattern, analogous to the warping of a scene that is viewed with visible light through an old uneven glass window. This is analogous to early visible-light experiments that photographed a grating pattern to pick up the refractive index variations introduced by a jet of gas or hot air placed between the camera and the reference pattern \cite{Massig1,Massig2,Perciante}. In this approach, the dark-field signal will be seen as a blurring of the grating pattern, similar to how a scene viewed through a ground glass window is blurred.  The x-ray phase and dark-field images are retrieved by either analyzing sample-induced changes to the grating pattern in Fourier space, in the case that the illumination is well-described by a sinusoid \cite{Wen2010, takeda1982}, or in real-space, via a series of local cross-correlations \cite{Morgan2011}. 

Because the real-space approach to phase and dark-field retrieval does not require a periodic reference pattern, a new set-up was proposed where a highly-textured random object like a piece of sandpaper could be used to create a speckle reference illumination \cite{berujon2012,morgan2012}.  The approach has come to be known in the x-ray regime as `speckle-tracking', and initially retrieved phase and dark-field images by comparing a reference speckle image without the sample to a single exposure where the sample has now been introduced. This speckle reference pattern, created by the local focusing or defocusing of the x-ray illumination by sandpaper grains, calls back to early work on x-ray wavefront characterization that analyzed a reference intensity pattern created by an array of x-ray lenses \cite{MayoSexton2004}, a form of Shack-Hartmann sensor. This approach, taking advantage of the speckle characteristics in the deep Fresnel region where the Fresnel number is on the order of unity, proved itself particularly advantageous in the x-ray regime where the range of usable propagation distances is elongated by the short photon wavelength \cite{cerbino2008}. Later, equivalent methods were developed with visible light where the so-called speckle-memory effect also permits tracking of individual grains between images \cite{berto2017}.  

While the speckle-based approach provides improved spatial resolution compared to the early x-ray lens approach, because any x-ray speckles will be spread across several detector pixels, the spatial resolution of the retrieved images will be limited with this single-sample-exposure approach. As a result, the speckle-tracking method has evolved to incorporate information from multiple exposures, captured as speckles are scanned across the sample \cite{BerujonWangSawhney2012}. Since the first demonstration \cite{BerujonWangSawhney2012}, a number of additional phase and dark-field retrieval methods have been shown that use multiple sample exposures \cite{berujon2015,Berujon2015c,wang2016,Zdora2017}. Another key step has been the demonstration of this technique on laboratory x-ray sources, in both the single \cite{Zanette2014, wang2016lab} and multiple-exposure approaches \cite{zhou2015speckle}. Several recent reviews on speckle-tracking provide a full picture of the developments \cite{zdora2018,berujon2020a,berujon2020b}.

Two major formalisms for x-ray speckle tracking are X-ray Speckle-Vector Tracking (XSVT) \cite{Berujon2015c} and Unified Modulated Pattern Analysis (UMPA) \cite{Zdora2017}.  Both have a broad domain of applicability, and incorporate multiple important factors in a robust manner, via a variational approach based on a suitable functional.  The incorporated factors include attenuation and refraction (transverse phase shifts) due to the sample.  In addition, XSVT and UMPA can incorporate and 
retrieve the `dark-field' speckle-visibility reduction associated with the position-dependent small-angle x-ray scattering (SAXS) \cite{GlatterKratky1982} fans, that emerge from each point on the exit-surface of the sample on account of spatially-unresolved microstructure \cite{Pagot2003,Rigon2003,Wernick2003,Khelashvili2005,Pfeiffer2008,Kitchen2010,BerujonWangSawhney2012,Endrizzi2014,Zanette2014}.  For many samples, these position-dependent SAXS fans may be modeled as each being rotationally-symmetric.  When the position-dependent SAXS fans are not rotationally symmetric, e.g.~if they can be modeled as elliptical, one can instead speak of `directional dark field' (DDF) imaging \cite{jensen2010a,jensen2010b,jud2017}.  The DDF signal may be accessed using methods employing periodic gratings \cite{jensen2010a,jensen2010b,jud2017} and spatially-random gratings \cite{WangPNAS2021}.%

More recently, another random-mask speckle-tracking approach was developed.  This third approach is the Optical Flow (OF) method  \cite{PaganinLabrietBrunBerujon2018}, together with its Fokker--Planck generalization (Multi-modal Intrinsic Speckle Tracking (MIST)) \cite{PaganinMorgan2019,MIST1}. 
We now focus attention on this third method, which implicitly rather than explicitly tracks speckles. The OF method is based on a simple second-order partial differential equation, namely a continuity equation that has strong parallels with the transport-of-intensity equation of paraxial optics \cite{teague1983}.  This simplicity is obtained at the cost of being significantly less general than the XSVT and UMPA approaches.  In the MIST extension of OF, a Fokker--Planck  \cite{Risken1989} generalization of OF speckle tracking is employed \cite{MorganPaganin2019,PaganinMorgan2019,MIST1}.  MIST implicitly tracks the transverse motion, lensing and diffusion of speckles.  Its associated partial differential equation \cite{PaganinMorgan2019}, which is of Fokker--Planck form, is amenable to closed-form solution \cite{MIST1}.  The latter fact is the core motivation for pursuing this particular approach.  

The paper is structured as follows.  Section II develops a theoretical description, for directional dark-field x-ray speckle tracking, via a forward-finite-difference anisotropic-diffusion Fokker--Planck equation.  The key aim, underpinning this theory, is robust means for extracting the position-dependent symmetric rank-two diffusion tensor associated with unresolved microstructure in a sample.  Sections III and IV give an experimental demonstration of these ideas, using hard x-ray radiation.  Section V discusses some broader implications of this work, together with some possible avenues for future investigation.  Brief concluding remarks are made in Sec.~VI.  

\section{Theory}

Here we develop the theory for both forward and inverse problems \cite{Sabatier2000} of directional dark-field x-ray speckle tracking, using an anisotropic-diffusion Fokker--Planck equation.  We begin by obtaining the requisite forward-finite-difference Fokker--Planck equation, which models speckle formation and subsequent sample-induced deformation, in a manner accounting for the attenuating, refracting and diffusive properties of the sample. We then consider the inverse problem associated with two different special cases of this model, namely (i) a phase object, for which there is no attenuation by assumption, and (ii) a monomorphous object, in which the object-induced phase shifts are proportional to the logarithm of the associated attenuation.  We then examine the relationship between the diffusion-tensor field appearing in our Fokker--Planck equation, and the associated position-dependent SAXS fans associated with unresolved microstructure in the sample.

\subsection{Anisotropic-diffusion Fokker--Planck formalism for x-ray speckle tracking}

\begin{figure}
\includegraphics[width=0.46\textwidth]{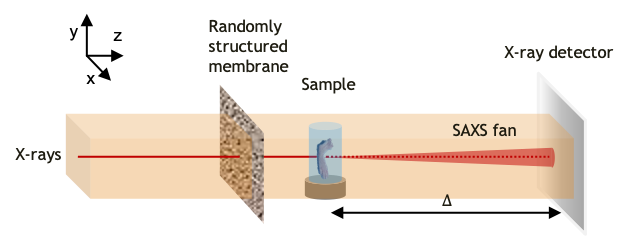}
\caption{Experimental setup for x-ray Multi-modal Instrinsic-Speckle-Tracking (MIST).\label{fig:ExperimentalSchematic}}
\end{figure}

Consider the x-ray speckle-tracking setup that is sketched in Fig.~\ref{fig:ExperimentalSchematic}.  Here, we see a quasi-monochromatic x-ray source that paraxially illuminates a thin spatially-random speckle-generating mask, to produce a reference speckle image $I_{\textrm R}(x,y)$, over the planar surface of a position-sensitive detector which is perpendicular to the optical axis $z$.  The reference speckle image may be produced via either or both of the following mechanisms: (i) attenuation contrast due to the position-dependent absorption of x rays as they traverse the spatially-random mask, (ii) phase contrast due to the position-dependent phase shifts imparted upon the x rays as they traverse the mask, with these phase shifts leading upon free-space propagation to intensity variations over the surface of the position-sensitive detector \cite{Snigirev1995}.  Having recorded the reference speckle image, we may then perform a second measurement, in which a thin non-absorbing object is placed in between the speckle-generating mask and the detector.  The distance $\Delta$ from the sample to the detector should be sufficiently large that the transverse deflections, induced by the refractive profile of the sample, lead to measurable transverse shifts in the reference speckles measured in the absence of the sample. Conversely, $\Delta$ should be sufficiently small that the transverse shifts, of the reference speckles, should not be more than around two detector pixels in magnitude.  The speckle image $I_{\textrm S}(x,y)$, measured in the presence of both the sample and the mask, will then have the following property: every x-ray photon that strikes a given location $(x,y)$ in $I_{\textrm S}$ will have struck a nearby position $(x+\delta x,y+\delta y)$ in $I_{\textrm R}$, and conversely.  This property, which arises from the previously articulated assumption that $\Delta$ be sufficiently small, implies $I_{\textrm R}$ and $I_{\textrm S}$ to be connected via a conserved current that locally preserves the number of detected photons.  Stated differently, local conservation of energy implies there to be a geometric flow---which may also be termed an optical flow \cite{OF_old1,OF_old2} or a Noether conserved current \cite{MandlShawBook}---that can be used to smoothly deform $I_{\textrm R}$ into $I_{\textrm S}$ \cite{PaganinLabrietBrunBerujon2018}.

It is natural to assume that this flow, which deforms $I_{\textrm R}$ into $I_{\textrm S}$, will have both coherent and diffuse components. The coherent component will be due to the near-monochromatic nature of the illuminating radiation, together with the position-dependent phase and amplitude shifts that are coherently imparted by both the sample and the speckle-generating mask.  The diffuse component will be due to several mechanisms: (i) the partially-coherent nature of the illumination, (ii) the presence of spatially-unresolved random micro-structure in the sample and mask, and (iii) the presence of sharp edges in both sample and mask.  Spatially-unresolved random micro-structure leads to diffusive x-ray energy flow on account of position-dependent small-angle x-ray scattering (SAXS) \cite{GlatterKratky1982}.  Sharp edges lead to diffusive flow via Young--Maggi--Rubinowicz boundary-diffraction waves (edge scattering, edge-diffracted rays) \cite{YoungOnTheBoundaryWave,Maggi,Rubinowicz,MiyamotoWolf1,MiyamotoWolf2}.    

All of the above considerations---namely a locally energy-preserving optical flow, possessing both coherent and diffuse components, that maps $I_{\textrm R}$ into $I_{\textrm S}$---may be quantified using a forward-finite-difference Fokker--Planck equation \cite{Risken1989,MorganPaganin2019,PaganinMorgan2019,PaganinPelliccia2021}.  In particular, Eq.~(55) in Ref.~\cite{PaganinMorgan2019} generalizes the optical-flow method for x-ray speckle tracking \cite{PaganinLabrietBrunBerujon2018}, to give:
\begin{align}
  \nonumber I_{\textrm R}({\bf{r}}_{\perp})-I_{\textrm S}({\bf{r}}_{\perp})= & \frac{\Delta}{k} \nabla_{\perp}\cdot [I_{\textrm R}({\bf{r}}_{\perp}) \nabla_{\perp}\phi_{\textrm ob}({\bf{r}}_{\perp})] \\ &- \Delta\nabla_{\perp}^2[D_{\textrm{eff}}({\bf{r}}_{\perp};\Delta)I_{\textrm R}({\bf{r}}_{\perp})].   \label{eq:1} 
\end{align}
 Here ${\bf{r}}_{\perp} \equiv (x,y)$, $I_{\textrm R}({\bf{r}}_{\perp})$ and $I_{\textrm S}({\bf{r}}_{\perp})$ are the intensities of a reference speckle image acquired in the absence and presence of a sample, respectively, $\Delta$ is the sample-to-detector distance, $k=2\pi / \lambda$ is the wavenumber of the x rays, $\lambda$ is the wavelength, $\phi_{\textrm{ob}}$ is the phase shift caused by the sample, and $\nabla_{\perp}=(\partial/\partial x,\partial/\partial y)$ denotes the transverse gradient operator.  The first term, on the right-hand side of Eq.~(\ref{eq:1}), quantifies the coherent energy flow that was described earlier.  An identical term appears in the transport-of-intensity equation (TIE)  \cite{teague1983}.  Indeed, the entire first line of Eq.~(\ref{eq:1}) is mathematically identical in form to a first-order finite-difference form of the TIE.  This correspondence exists because local energy conservation, under transverse energy flow, is the underpinning principle in both the TIE and the first line of Eq.~(\ref{eq:1}).  The difference lies in the fact that (i) the transverse flow in the TIE is induced by free-space propagation from plane to parallel downstream plane {\em in vacuo}, whereas (ii) in the first line of Eq.~(\ref{eq:1}) the transverse flow occurs in a fixed plane, with the flow induced by a phase object whose presence smoothly perturbs $I_{\textrm R}({\bf{r}}_{\perp})$ into $I_{\textrm S}({\bf{r}}_{\perp})$ \cite{PaganinLabrietBrunBerujon2018}.  The second term, on the right-hand side of Eq.~(\ref{eq:1}), describes the diffusive flow via the effective diffusion coefficient $D_{\textrm{eff}}({\bf{r}}_{\perp};\Delta)$.  If we assume the mask to be spatially statistically stationary, and the source to be an extended incoherent source, then the diffusive effects due to source-size blur will be describable via a position-independent additive constant term in $D_{\textrm{eff}}({\bf{r}}_{\perp};\Delta)$.  With this understanding in place, we henceforth consider $D_{\textrm{eff}}({\bf{r}}_{\perp};\Delta)$ to quantify only sample-induced contributions, that are due to both (i) local position-dependent SAXS fans emerging from each point on the exit surface of the sample, as well as (ii) edge-diffracted x rays.  Note, also, that the position-dependent sample-induced SAXS fans, as described by the above formalism, are considered to be rotationally symmetric (by assumption) at each transverse location over the exit surface of the thin sample.
 
 To generalize Eq.~(\ref{eq:1}) to the case of directional dark-field imaging \cite{jensen2010a,jensen2010b}, in which the position-dependent SAXS fans have an elliptical transverse profile \cite{jensen2010b,yashiro2011}, we introduce the symmetric rank-two diffusion tensor field \cite{MorganPaganin2019,PaganinMorgan2019} :  
\begin{equation}
D_{\textrm{eff}}({\bf{r}}_{\perp};\Delta)\longrightarrow \begin{bmatrix}
D^{(xx)}_{\textrm{eff}}({\bf{r}}_{\perp};\Delta) & 
\tfrac{1}{2}D^{(xy)}_{\textrm{eff}}({\bf{r}}_{\perp};\Delta) \\
\tfrac{1}{2}D^{(xy)}_{\textrm{eff}}({\bf{r}}_{\perp};\Delta) & D^{(yy)}_{\textrm{eff}}({\bf{r}}_{\perp};\Delta)
\end{bmatrix} . 
\label{eq:DiffusionTensor}
\end{equation}
This enables us to write down a directional-dark-field generalization of Eq.~(\ref{eq:1}), namely the following anisotropic-diffusion forward-finite-difference Fokker--Planck speckle-tracking equation due to \citet{MIST1}:  
\begin{align}
  \nonumber I_{\textrm R}({\bf{r}}_{\perp}) -& I_{\textrm S}({\bf{r}}_{\perp}) = \frac{\Delta}{k} \nabla_{\perp}\cdot [I_{\textrm R}({\bf{r}}_{\perp}) \nabla_{\perp}\phi_{\textrm ob}({\bf{r}}_{\perp})] \\ \nonumber &- \Delta\frac{\partial^2}{\partial x^2}\left[D^{(xx)}_{\textrm{eff}}({\bf{r}}_{\perp};\Delta)I_{\textrm R}({\bf{r}}_{\perp})\right]  
  \\ \nonumber &- \Delta\frac{\partial^2}{\partial y^2}\left[D^{(yy)}_{\textrm{eff}}({\bf{r}}_{\perp};\Delta)I_{\textrm R}({\bf{r}}_{\perp})\right]  
  \\  &- \Delta\frac{\partial^2}{\partial x\partial y}\left[D^{(xy)}_{\textrm{eff}}({\bf{r}}_{\perp};\Delta)I_{\textrm R}({\bf{r}}_{\perp})\right].
  \label{eq:1--generalised-form} 
\end{align}

Let us now assume that an attenuating object is placed in the well-resolved reference speckle field. This causes variations in the registered speckle image, which can be described by (cf.~Eq.~(9) in Ref.~\cite{PavlovPhysRevAppl2020}, with incorporation of Eq.~(4) in Ref.~\cite{MorganPaganin2019} and Eq.~(51) in Ref.~\cite{PaganinMorgan2019}):%
\begin{align}
  \nonumber I_{\textrm R}({\bf{r}}_{\perp}) & I_{\textrm ob}({\bf{r}}_{\perp}) - I_{\textrm S}({\bf{r}}_{\perp}) \\ \nonumber =& \frac{\Delta}{k} \nabla_{\perp}\cdot [I_{\textrm R}({\bf{r}}_{\perp})I_{\textrm ob}({\bf{r}}_{\perp}) \nabla_{\perp}\phi_{\textrm ob}({\bf{r}}_{\perp})]
  \quad \\ \nonumber &- \Delta\frac{\partial^2}{\partial x^2}\left[D^{(xx)}_{\textrm{eff}}({\bf{r}}_{\perp};\Delta)I_{\textrm R}({\bf{r}}_{\perp})I_{\textrm ob}({\bf{r}}_{\perp})\right]  
  \\ \nonumber &- \Delta\frac{\partial^2}{\partial y^2}\left[D^{(yy)}_{\textrm{eff}}({\bf{r}}_{\perp};\Delta)I_{\textrm R}({\bf{r}}_{\perp})I_{\textrm ob}({\bf{r}}_{\perp})\right]  
  \\  &- \Delta\frac{\partial^2}{\partial x\partial y}\left[D^{(xy)}_{\textrm{eff}}({\bf{r}}_{\perp};\Delta)I_{\textrm R}({\bf{r}}_{\perp})I_{\textrm ob}({\bf{r}}_{\perp})\right].
  \label{starting form} 
\end{align}
Here $I_{\textrm ob}({\bf{r}}_{\perp})$ is the object's attenuation term. We also assume that the components of the diffusion tensor, $D^{(xx,yy,xy)}_{\textrm{eff}}$, are  slowly-varying functions of the transverse position (i.e., we can neglect their transverse spatial derivatives, which are small compared to the retained terms).

The second-order transverse spatial derivatives, applied to the diffuse-scatter terms on the right side of Eq.~(\ref{starting form}), yield several components.  For example, the first of these three diffusely-scattering components gives:%
\begin{align} \nonumber \frac{\partial^2}{\partial x^2}  & \left[D^{(xx)}_{\textrm{eff}}  ({\bf{r}}_{\perp};\Delta)I_{\textrm R}({\bf{r}}_{\perp})I_{\textrm ob}({\bf{r}}_{\perp})\right] 
 \\ \nonumber =& \left[D^{(xx)}_{\textrm{eff}}({\bf{r}}_{\perp};\Delta)I_{\textrm ob}({\bf{r}}_{\perp})\right]\frac{\partial^2}{\partial x^2}I_{\textrm R}({\bf{r}}_{\perp}) 
  \\ \nonumber &+ I_{\textrm R}({\bf{r}}_{\perp})
  \frac{\partial^2}{\partial x^2}\left[D^{(xx)}_{\textrm{eff}}({\bf{r}}_{\perp};\Delta)I_{\textrm ob}({\bf{r}}_{\perp})\right]  
  \\ \nonumber &+ 2\frac{\partial}{\partial x}[I_{\textrm R}({\bf{r}}_{\perp})]
  \frac{\partial}{\partial x}\left[D^{(xx)}_{\textrm{eff}}({\bf{r}}_{\perp};\Delta)I_{\textrm ob}({\bf{r}}_{\perp})\right] 
  \\ \nonumber \approx & D^{(xx)}_{\textrm{eff}}({\bf{r}}_{\perp};\Delta)I_{\textrm ob}({\bf{r}}_{\perp})
  \frac{\partial^2}{\partial x^2}I_{\textrm R}({\bf{r}}_{\perp}) 
  \\ &+ I_{\textrm R}({\bf{r}}_{\perp})D^{(xx)}_{\textrm{eff}}({\bf{r}}_{\perp};\Delta)
  \frac{\partial^2}{\partial x^2}[I_{\textrm ob}({\bf{r}}_{\perp})] 
  . 
  \label{details of second-derivatives} 
\end{align}
We have neglected the following terms in the right hand side of Eq.~(\ref{details of second-derivatives}): $I_{\textrm R}({\bf{r}}_{\perp})I_{\textrm ob}({\bf{r}}_{\perp})
  \frac{\partial^2}{\partial x^2}D^{(xx)}_{\textrm{eff}}({\bf{r}}_{\perp};\Delta) $ and $2I_{\textrm R}({\bf{r}}_{\perp})\frac{\partial}{\partial x}[D^{(xx)}_{\textrm{eff}}({\bf{r}}_{\perp};\Delta)]
  \frac{\partial}{\partial x}[I_{\textrm ob}({\bf{r}}_{\perp})]$, because we have assumed that $D^{(xx)}_{\textrm{eff}}$ is a slowly-varying function everywhere. We have also neglected the term $2\frac{\partial}{\partial x}[D^{(xx)}_{\textrm{eff}}({\bf{r}}_{\perp};\Delta)I_{\textrm ob}({\bf{r}}_{\perp})]
  \frac{\partial}{\partial x}[I_{\textrm R}({\bf{r}}_{\perp})]$, for the following reason. The intensity of a reference speckle image $I_{\textrm R}({\bf{r}}_{\perp})$, acquired in the absence of a sample, is produced by a spatially random mask. Therefore, the gradient of such an intensity map will be a vector field that is rapidly changing in both direction and magnitude. Thus the scalar product, of such a random vector field with a more slowly changing gradient of the product of two functions, can be neglected. A similar approximation was previously employed in Refs.~\cite{PavlovPhysRevAppl2020,MIST1}.
  Bearing all of the above points in mind, we can simplify Eq.~(\ref{starting form}) as follows:
\begin{align}
  \nonumber I_{\textrm R}({\bf{r}}_{\perp}) & I_{\textrm ob}({\bf{r}}_{\perp}) - I_{\textrm S}({\bf{r}}_{\perp})
  \\ \nonumber = & \frac{\Delta}{k}   I_{\textrm R}({\bf{r}}_{\perp}) \nabla_{\perp}\cdot [I_{\textrm ob}({\bf{r}}_{\perp}) \nabla_{\perp}\phi_{\textrm ob}({\bf{r}}_{\perp})]
  \\ \nonumber  &- \Delta D^{(xx)}_{\textrm{eff}}({\bf{r}}_{\perp};\Delta)I_{\textrm ob}({\bf{r}}_{\perp})
  \frac{\partial^2}{\partial x^2}I_{\textrm R}({\bf{r}}_{\perp}) 
  \\ \nonumber &- \Delta D^{(xx)}_{\textrm{eff}}({\bf{r}}_{\perp};\Delta)I_{\textrm R}({\bf{r}}_{\perp})
  \frac{\partial^2}{\partial x^2}I_{\textrm ob}({\bf{r}}_{\perp})
    \\ \nonumber  &- \Delta D^{(yy)}_{\textrm{eff}}({\bf{r}}_{\perp};\Delta)I_{\textrm ob}({\bf{r}}_{\perp})
  \frac{\partial^2}{\partial y^2}I_{\textrm R}({\bf{r}}_{\perp}) 
  \\ \nonumber &- \Delta D^{(yy)}_{\textrm{eff}}({\bf{r}}_{\perp};\Delta)I_{\textrm R}({\bf{r}}_{\perp})
  \frac{\partial^2}{\partial y^2}I_{\textrm ob}({\bf{r}}_{\perp})
    \\ \nonumber  &- \Delta D^{(xy)}_{\textrm{eff}}({\bf{r}}_{\perp};\Delta)I_{\textrm ob}({\bf{r}}_{\perp})
  \frac{\partial^2}{\partial x \partial y}I_{\textrm R}({\bf{r}}_{\perp}) 
  \\  &- \Delta D^{(xy)}_{\textrm{eff}}({\bf{r}}_{\perp};\Delta)I_{\textrm R}({\bf{r}}_{\perp})
  \frac{\partial^2}{\partial x \partial y}I_{\textrm ob}({\bf{r}}_{\perp}). 
  \label{starting form simplified} 
\end{align}
Above, we have used an additional approximation, similar to that used earlier, namely the neglect of the term $ \nabla_{\perp}[I_{\textrm R}({\bf{r}}_{\perp})]\cdot [I_{\textrm ob}({\bf{r}}_{\perp}) \nabla_{\perp}\phi_{\textrm ob}({\bf{r}}_{\perp})]$  (see also Refs.~\cite{PavlovPhysRevAppl2020,MIST1}). Here we are again making use of the fact that the intensity of a reference speckle image, acquired in the absence of a sample, $I_{\textrm R}$, is produced by a random mask. Therefore, the gradient of such a field is again a vector field that is rapidly changing in both direction and magnitude. Thus, the scalar product of such a random vector field with a more slowly changing phase gradient can be neglected. It is also worth noting that the terms on the right side of Eq.~(\ref{starting form simplified}), containing the second-order derivatives of the object’s attenuation term, become more prominent at the object’s internal and external boundaries. 

A further modification of of Eq.~(\ref{starting form simplified}) can be achieved by dividing both sides by $I_{\textrm R}$ and rearranging, to give:
\begin{align}
  \nonumber &\frac{I_{\textrm S}({\bf{r}}_{\perp})}{I_{\textrm R}({\bf{r}}_{\perp})} = I_{\textrm ob}({\bf{r}}_{\perp}) - \frac{\Delta}{k} \nabla_{\perp}\cdot [I_{\textrm ob}({\bf{r}}_{\perp}) \nabla_{\perp}\phi_{\textrm ob}({\bf{r}}_{\perp})]
  \\ \nonumber  &+ \Delta D^{(xx)}_{\textrm{eff}}({\bf{r}}_{\perp};\Delta) \left[I_{\textrm ob}({\bf{r}}_{\perp})
  \frac{ \frac{\partial^2}{\partial x^2}I_{\textrm R}({\bf{r}}_{\perp})}{I_{\textrm R}({\bf{r}}_{\perp})} + \frac{\partial^2}{\partial x^2}I_{\textrm ob}({\bf{r}}_{\perp})\right]
  \\ \nonumber  &+ \Delta D^{(yy)}_{\textrm{eff}}({\bf{r}}_{\perp};\Delta) \left[I_{\textrm ob}({\bf{r}}_{\perp})
  \frac{ \frac{\partial^2}{\partial y^2}I_{\textrm R}({\bf{r}}_{\perp})}{I_{\textrm R}({\bf{r}}_{\perp})} + \frac{\partial^2}{\partial y^2}I_{\textrm ob}({\bf{r}}_{\perp})\right]
  \\ &+ \Delta D^{(xy)}_{\textrm{eff}}({\bf{r}}_{\perp};\Delta) \left[I_{\textrm ob}({\bf{r}}_{\perp})
  \frac{ \frac{\partial^2}{\partial x \partial y}I_{\textrm R}({\bf{r}}_{\perp})}{I_{\textrm R}({\bf{r}}_{\perp})} + \frac{\partial^2}{\partial x \partial y}I_{\textrm ob}({\bf{r}}_{\perp})\right]
  . 
  \label{starting form simplified final} 
\end{align}
This equation describes a general case of speckle-based imaging for an attenuating object, where several terms incorporating the diffusion tensor field are taken into account.  This forward-finite-difference anisotropic-diffusion Fokker--Planck equation completes our description of the forward problem associated with image formation and subsequent data collection, in the context of x-ray speckle tracking. 

\subsection{Two inverse problems}

The formulation of the forward problem in Eq.~(\ref{starting form simplified final}) establishes the model upon which the associated inverse problem \cite{Sabatier2000}, of recovering sample properties based on one or more pairs of reference-only and reference-plus-sample speckle images, may be based.  Here we consider two such inverse problems, corresponding to two different limit cases of Eq.~(\ref{starting form simplified final}).  These two special cases are: (i) the sample is assumed to be a phase object, i.e.~it is non-attenuating; (ii) the sample is taken to be monomorphous, i.e.~its phase shifts are proportional to the logarithm of the associated attenuation, as is the case e.g.~for a thin object that is composed of a single material. 

\subsubsection{Case 1: A Phase object}

By definition, $I_{\textrm ob}({\bf{r}}_{\perp})=1$ for a pure phase object. Equation (\ref{starting form simplified final}) can then be simplified as (cf.~Eq.~(10) in Ref.~\cite{MIST1} and to Eq.~(\ref{eq:1--generalised-form}) above):%
\begin{align}
  \nonumber 1-\frac{I_{\textrm S}({\bf{r}}_{\perp})}{I_{\textrm R}({\bf{r}}_{\perp})} =& \frac{\Delta}{k} \nabla^2_{\perp} \phi_{\textrm ob}({\bf{r}}_{\perp})
  \\ \nonumber &- \Delta D^{(xx)}_{\textrm{eff}}({\bf{r}}_{\perp};\Delta) 
  \frac{ \frac{\partial^2}{\partial x^2}I_{\textrm R}({\bf{r}}_{\perp})}{I_{\textrm R}({\bf{r}}_{\perp})} 
  \\ \nonumber   &- \Delta D^{(yy)}_{\textrm{eff}}({\bf{r}}_{\perp};\Delta) 
  \frac{ \frac{\partial^2}{\partial y^2}I_{\textrm R}({\bf{r}}_{\perp})}{I_{\textrm R}({\bf{r}}_{\perp})} 
  \\ &- \Delta D^{(xy)}_{\textrm{eff}}({\bf{r}}_{\perp};\Delta) 
  \frac{ \frac{\partial^2}{\partial x \partial y}I_{\textrm R}({\bf{r}}_{\perp})}{I_{\textrm R}({\bf{r}}_{\perp})}
  . 
  \label{phase object} 
\end{align}
Equation ~(\ref{phase object}) contains four unknown functions: $\nabla^2_{\perp} \phi({\bf{r}}_{\perp})$ and $D^{(xx,xy,yy)}_{\textrm{eff}}({\bf{r}}_{\perp};\Delta)$,  which can be found from any four independent measurements of  $I_{\textrm S}({\bf{r}}_{\perp})$ and $I_{\textrm R}({\bf{r}}_{\perp})$  by varying experimental conditions, e.g., the mask positions.

\subsubsection{Case 2: A monomorphous object}

For a monomorphous (e.g., single-material) object, its complex index of refraction $n$ can be represented as
\begin{eqnarray}
\label{refraction index single material}
n=1-\delta+i\beta=1-\gamma\beta+i\beta=1+\beta(i-\gamma),
\end{eqnarray}
where 
\begin{equation}
\label{definition of gamma}
\gamma=\frac{\delta}{\beta}.    
\end{equation}
Here, the real numbers $\delta,\beta$ denote the refractive index decrement and the imaginary (absorptive) part of the complex refractive index, respectively. The value of $\gamma$ is considered known from tables or can be experimentally adjusted by trial and error to match the sample composition and density (see e.g., Ref.~\cite{delRio2011}).
Then the phase shift is
\begin{equation}
\label{definition of phase of monomorphous objects}
\phi({\bf{r}}_{\perp})=-k \delta t({\bf{r}}_{\perp}),    
\end{equation}
where $t({\bf{r}}_{\perp})$ is the projected thickness of the object. The object's attenuation term is
\begin{equation}
\label{definition of object's attenuation term of monomorphous objects}
I_{\textrm ob}({\bf{r}}_{\perp})=\exp[-2k \beta t({\bf{r}}_{\perp})].
\end{equation}
Then Eq.~(\ref{starting form simplified final}) can be rewritten as follows (cf.~Eq.~(14) in Ref.~\cite{PavlovPhysRevAppl2020}):%
\begin{align}
  \nonumber & \frac{I_{\textrm S}({\bf{r}}_{\perp})}{I_{\textrm R}({\bf{r}}_{\perp})} = \left(1-\frac{\gamma \Delta}{2k}\nabla^2_{\perp}\right)I_{\textrm ob}({\bf{r}}_{\perp}) \\ \nonumber  &+ \Delta D^{(xx)}_{\textrm{eff}}({\bf{r}}_{\perp};\Delta) \left[I_{\textrm ob}({\bf{r}}_{\perp})
  \frac{ \frac{\partial^2}{\partial x^2}I_{\textrm R}({\bf{r}}_{\perp})}{I_{\textrm R}({\bf{r}}_{\perp})} + \frac{\partial^2}{\partial x^2}I_{\textrm ob}({\bf{r}}_{\perp})\right]
  \\ \nonumber  &+ \Delta D^{(yy)}_{\textrm{eff}}({\bf{r}}_{\perp};\Delta) \left[I_{\textrm ob}({\bf{r}}_{\perp})
  \frac{ \frac{\partial^2}{\partial y^2}I_{\textrm R}({\bf{r}}_{\perp})}{I_{\textrm R}({\bf{r}}_{\perp})} + \frac{\partial^2}{\partial y^2}I_{\textrm ob}({\bf{r}}_{\perp})\right]
  \\ \nonumber  &+ \Delta D^{(xy)}_{\textrm{eff}}({\bf{r}}_{\perp};\Delta) \left[I_{\textrm ob}({\bf{r}}_{\perp})
  \frac{ \frac{\partial^2}{\partial x \partial y}I_{\textrm R}({\bf{r}}_{\perp})}{I_{\textrm R}({\bf{r}}_{\perp})} + \frac{\partial^2}{\partial x \partial y}I_{\textrm ob}({\bf{r}}_{\perp})\right]
  . \\
  \label{starting form simplified final for monomorphous obects} 
\end{align}
As we assume that $D^{(xx,yy,xy)}_{\textrm{eff}}({\bf{r}}_{\perp};\Delta)$ are slowly-varying functions, we can solve Eq.~(\ref{starting form simplified final for monomorphous obects}) for four unknown functions, namely:%
\begin{align}
  \nonumber G_{\textrm 1}({\bf{r}}_{\perp}) =& \left(1-\frac{\gamma \Delta}{2k}\nabla^2_{\perp}\right)I_{\textrm ob}({\bf{r}}_{\perp})
  \\ \nonumber &+ \Delta\left[D^{(xx)}_{\textrm{eff}}({\bf{r}}_{\perp};\Delta)\frac{\partial^2}{\partial x^2}I_{\textrm ob}({\bf{r}}_{\perp})\right]
  \\ \nonumber &+ \Delta\left[D^{(xy)}_{\textrm{eff}}({\bf{r}}_{\perp};\Delta)\frac{\partial^2}{\partial x \partial y}I_{\textrm ob}({\bf{r}}_{\perp})\right]
  \\  &+ \Delta\left[D^{(yy)}_{\textrm{eff}}({\bf{r}}_{\perp};\Delta)\frac{\partial^2}{\partial y^2}I_{\textrm ob}({\bf{r}}_{\perp})\right]
  ,\label{four unknown functions No 1} 
\end{align}
\begin{eqnarray}
  G_{\textrm 2}({\bf{r}}_{\perp}) = 
   \Delta\left[D^{(xx)}_{\textrm{eff}}({\bf{r}}_{\perp};\Delta)I_{\textrm ob}({\bf{r}}_{\perp})\right]
  , 
  \label{four unknown functions No 2} 
\end{eqnarray}
\begin{eqnarray}
  G_{\textrm 3}({\bf{r}}_{\perp}) = 
   \Delta\left[D^{(yy)}_{\textrm{eff}}({\bf{r}}_{\perp};\Delta)I_{\textrm ob}({\bf{r}}_{\perp})\right]
  , 
  \label{four unknown functions No 3} 
\end{eqnarray}
\begin{eqnarray}
  G_{\textrm 4}({\bf{r}}_{\perp}) = 
   \Delta\left[D^{(xy)}_{\textrm{eff}}({\bf{r}}_{\perp};\Delta)I_{\textrm ob}({\bf{r}}_{\perp})\right]
  . 
  \label{four unknown functions No 4} 
\end{eqnarray}
This can be done by using four independent measurements of  $I_{\textrm S}({\bf{r}}_{\perp})$ and $I_{\textrm R}({\bf{r}}_{\perp})$, which may be obtained by varying experimental conditions, e.g., the mask positions. Then we can apply the second-order derivatives to the functions $G_{\textrm 2,3,4}({\bf{r}}_{\perp})$:

\begin{eqnarray}\label{three unknown functions 2nd derivatives} 
  \begin{cases}
    \frac{\partial^2}{\partial x^2}G_{\textrm 2}({\bf{r}}_{\perp}) \approx
   \Delta D^{(xx)}_{\textrm{eff}}({\bf{r}}_{\perp};\Delta)\frac{\partial^2}{\partial x^2}I_{\textrm ob}({\bf{r}}_{\perp}) \\
    \frac{\partial^2}{\partial y^2}G_{\textrm 3}({\bf{r}}_{\perp}) \approx 
   \Delta D^{(yy)}_{\textrm{eff}}({\bf{r}}_{\perp};\Delta)\frac{\partial^2}{\partial y^2}I_{\textrm ob}({\bf{r}}_{\perp})
    \\
    \frac{\partial^2}{\partial x \partial y}G_{\textrm 4}({\bf{r}}_{\perp}) \approx
   \Delta D^{(xy)}_{\textrm{eff}}({\bf{r}}_{\perp};\Delta)\frac{\partial^2}{\partial x \partial y}I_{\textrm ob}({\bf{r}}_{\perp})
    .  
    \end{cases}
\end{eqnarray}
Here we have again used our assumption that $D^{(xx,yy,xy)}_{\textrm{eff}}({\bf{r}}_{\perp};\Delta)$ are slowly-varying functions. By combining Eq.~(\ref{four unknown functions No 1}) and Eq.~(\ref{three unknown functions 2nd derivatives}) we can form a new function: 
\begin{eqnarray}
  \nonumber G({\bf{r}}_{\perp}) 
  = G_{\textrm 1}({\bf{r}}_{\perp}) - \frac{\partial^2}{\partial x^2}G_{\textrm 2}({\bf{r}}_{\perp}) -\frac{\partial^2}{\partial y^2}G_{\textrm 3}({\bf{r}}_{\perp}) 
  \\ \nonumber  -\frac{\partial^2}{\partial x \partial y}G_{\textrm 4}({\bf{r}}_{\perp}) 
  = \left(1-\frac{\gamma \Delta}{2k}\nabla^2_{\perp}\right)I_{\textrm ob}({\bf{r}}_{\perp})
  . \\
  \label{unknown function G} 
\end{eqnarray}
Now we can obtain the object's projected thickness map (cf.~Eq.~(18) in Ref.~\cite{PavlovPhysRevAppl2020}):
\begin{eqnarray}
\label{calculated projected thickness}
t({\bf{r}}_{\perp}) = -\frac{1}{2k\beta} \, \log_e\mathcal{F}^{-1} \left\{ \frac{\mathcal{F}\left[G({\bf{r}}_{\perp})\right]}{1+\pi\gamma \Delta \lambda (u^2+v^2)} \right\} 
\end{eqnarray}
and subsequently the components of the diffusion tensor:
\begin{eqnarray}\label{diffusion tensor components} 
  \begin{cases}
    D^{(xx)}_{\textrm{eff}}({\bf{r}}_{\perp};\Delta) = G_{\textrm 2}({\bf{r}}_{\perp})/(\Delta I_{\textrm ob}({\bf{r}}_{\perp}))
    \\
    D^{(yy)}_{\textrm{eff}}({\bf{r}}_{\perp};\Delta) = G_{\textrm 3}({\bf{r}}_{\perp})/(\Delta I_{\textrm ob}({\bf{r}}_{\perp}))
    \\
    D^{(xy)}_{\textrm{eff}}({\bf{r}}_{\perp};\Delta) = G_{\textrm 4}({\bf{r}}_{\perp})/(\Delta I_{\textrm ob}({\bf{r}}_{\perp}))
    .  
    \end{cases}
\end{eqnarray}

\subsection{Relation  between  diffusion-tensor field and position-dependent SAXS fans}

Here we consider the relationship between the diffusion-tensor field (Eq.~(\ref{eq:DiffusionTensor})) and the associated position-dependent elliptical SAXS fans emanating from each point on the exit surface of the sample.  

If we consider the transverse location  $(x_0,y_0)$ at the nominally-planar exit surface of the sample, then the resulting anisotropic blurring due to the locally-elliptical SAXS fan will correspond to an ellipse of $(x,y)$ coordinates in the plane at distance $\Delta > 0$ downstream of the sample, with these $(x,y)$ coordinates obeying:   
\begin{align}
 \nonumber \frac{(x-x_0)^2}{D^{(xx)}_{\textrm{eff}}(x_0,y_0;\Delta)}+\frac{(y-y_0)^2}{D^{(yy)}_{\textrm{eff}}(x_0,y_0;\Delta)} \\ +  \frac{(x-x_0)(y-y_0)}{D^{(xy)}_{\textrm{eff}}(x_0,y_0;\Delta)}  \le \Delta.
 \label{eq:FamilyOfSAXSellipses}
\end{align}

The above ellipse field, in which we have a different ellipse for each transverse location $(x_0,y_0)$, incorporates the effects of two distinct but related factors: (i) the opening angles $\theta(x_0,y_0)$ of the elliptical SAXS fans that emanate from each point $(x_0,y_0)$ over the nominally-planar exit surface of the sample; (ii) the corresponding dimensionless fractions $F(x_0,y_0)$ of the incident x-ray photons that are converted to the diffusely-scattering channel, on account of unresolved micro-structure and edge scatter from the sample.  

See Fig.~3 of Paganin \& Morgan \cite{PaganinMorgan2019} for the relations connecting the following quantities: (i) the effective diffusion coefficients, (ii) the position-dependent SAXS-fan opening angles, and (iii) the diffuse-scatter fractions.  For an example of such connecting relations, for the `$xx$' diffuse-scatter channel we have 
\begin{equation}
 \frac{D^{(xx)}_{\textrm{eff}}(x_0,y_0;\Delta)}{\Delta}=F^{(xx)}(x_0,y_0)\left[\theta^{(xx)}(x_0,y_0)\right]^2.  
 \label{eq:RelatingDandFandTheta}
 \end{equation}
Here, $F^{(xx)}(x_0,y_0)$ is a dimensionless quantity taking values between zero and unity, that denotes the fraction of the incident x rays (at the specified energy) that are converted to `$xx$' diffuse scatter by the sample at the transverse location $(x_0,y_0)$, with $\theta^{(xx)}(x_0,y_0)$ being the corresponding local-SAXS-fan opening angle.  Similar expressions may be written down, by replacing $(xx)$ with either $(yy)$ or $(xy)$.  Note, also, that these three opening angles may be converted to (i) an angle for the SAXS-ellipse semi-major axis, (ii) an angle for the SAXS-ellipse semi-minor axis, and (iii) an angle denoting the angular orientation of the semi-major ellipse axis.   

The form of the right side of Eq.~(\ref{eq:RelatingDandFandTheta}) implies that the effective diffusion coefficients are invariant under the concurrent mappings 
\begin{equation}
    \begin{cases}
      F(x_0,y_0) \rightarrow F(x_0,y_0) / \alpha(x_0,y_0) , \\
      \theta(x_0,y_0)\rightarrow  \theta(x_0,y_0) \sqrt{\alpha(x_0,y_0)}.
    \end{cases}
\label{eq:Invariance}
\end{equation}
Here, $\alpha(x_0,y_0)$ is a real positive function which may assume otherwise-arbitrary values, provided that, both before and after the above mapping, the scattering fractions obey $F(x_0,y_0) \ll 1$ and the scattering angles obey $\theta(x_0,y_0) \ll 1$.  The physical origin of this invariance is as follows.  Decreasing $F(x_0,y_0)$ at any fixed transverse location will decrease the degree of local-SAXS blurring in the detection plane.  Conversely, increasing the corresponding SAXS-fan opening angle at the same sample location will increase the degree of such blurring.  These two opposing influences can be chosen to exactly balance one another (in the sense of providing an identical measured map of radiant exposure), in a continuous infinity of different ways, corresponding to all of the different choices that can be made for the scalar field $\alpha(x_0,y_0)$.  This is a fundamental ambiguity in the Fokker--Planck formalism that underpins the present paper.

The above ambiguity implies that, rather than recovering the SAXS-fan ellipse at any transverse location $(x_0,y_0)$, the method is only able to recover a family of similar concentric ellipses at each $(x_0,y_0)$ location.  Each member of this family is similar to the actual SAXS ellipse at each transverse location $(x_0,y_0)$, in the sense of having the same eccentricity and orientation angle, but which member of the family is the actual SAXS ellipse remains undetermined.   

Fortunately, there are two absolute quantities that may be extracted, since they are both independent of $\alpha(x_0,y_0)$ and are therefore the same for all ellipses in the previously mentioned family of similar ellipses.  These two invariant directional-dark-field quantities are:
\begin{itemize}
    \item the eccentricity $\epsilon(x_0,y_0)$ of the SAXS ellipse at each transverse location $(x_0,y_0)$;
    \item the angular orientation $\psi(x_0,y_0)$ of each SAXS ellipse (note that these angular orientations are only meaningful modulo $\pi$ radians, since the major axes of the ellipses form a director field rather than a vector field, i.e.~they are `arrowless vectors').
\end{itemize}  
Both $\epsilon(x_0,y_0)$ and $\psi(x_0,y_0)$ (positive direction is counterclockwise) may be extracted from the symmetrical quadratic form (see Eq. 2.4-1 in \cite{korn_korn_1968}) of an ellipse, where we assume that the center of this ellipse is at $(x_0,y_0)$:
\begin{equation}
 a_{11}x^{2}+2a_{12}xy+a_{22}y^{2}+a_{33} = 0.
  \label{eq:QuadraticForm}
 \end{equation}
Hence
\begin{equation}
    \begin{cases}
    \epsilon(x_0,y_0) = 
    \sqrt{2 \Upsilon/(a_{11}+a_{22}+\Upsilon)} \\ 
    \psi(x_0,y_0) = \frac{1}{2} \arctan[2a_{12}/(a_{22}-a_{11})] ,
    \end{cases}
\label{eq:eccentricity_angle}
\end{equation}
where 
\begin{equation}
\Upsilon=\sqrt{(a_{11}-a_{22})^2+4a_{12}^2},
\end{equation}
which corresponds to Eq.~(\ref{eq:FamilyOfSAXSellipses}).

Subsequently, using Eq.~(\ref{eq:eccentricity_angle}), both $\epsilon(x_0,y_0)$ and $\psi(x_0,y_0)$ may be extracted directly from the effective diffusion tensor in Eq.~(\ref{eq:DiffusionTensor}), via the following relations: 
\begin{eqnarray}\label{a11_a22_a12_Dxx_Dyy_Dxy} 
  \begin{cases}
    a_{11} = D^{(yy)}_{\textrm{eff}}({\bf{r}}_{\perp};\Delta)D^{(xy)}_{\textrm{eff}}({\bf{r}}_{\perp};\Delta)
    \\
    a_{22} = D^{(xx)}_{\textrm{eff}}({\bf{r}}_{\perp};\Delta)D^{(xy)}_{\textrm{eff}}({\bf{r}}_{\perp};\Delta)
    \\
    a_{12} = \frac{1}{2} D^{(xx)}_{\textrm{eff}}({\bf{r}}_{\perp};\Delta)D^{(yy)}_{\textrm{eff}}({\bf{r}}_{\perp};\Delta)
    .  
    \end{cases}
\end{eqnarray}
Thus, while this symmetric rank-two diffusion tensor contains three independent components at each transverse location $(x_0,y_0)$, the invariance under the mapping of Eq.~(\ref{eq:Invariance}) implies that only two independent invariant quantities may be extracted using our Fokker--Planck speckle-tracking formalism.  

Evidently, we have four independent recoverable channels of information in total, corresponding to four different scalar fields  at the exit surface of the sample. The first pair of scalar fields, which is associated with the coherent channel for x-ray energy flow, is the intensity $I_{\textrm ob}(x_0,y_0)$ and phase $\phi_{\textrm ob}(x_0,y_0)$ at each point $(x_0,y_0)$ on the exit surface of the sample.  The other pair of scalar fields, which is  associated with the diffuse channel for x-ray energy flow, is the eccentricity $\epsilon(x_0,y_0)$ and ellipse-orientation $\psi(x_0,y_0)$ of the local position-dependent SAXS ellipses emerging from each point on the sample's  exit surface.  A fifth scalar field, namely the dimensionless scattering fraction $F(x_0,y_0)$, is not recovered by the formalism developed in the present paper. 

The recovered ellipse-eccentricity and ellipse-orientation fields, namely $\epsilon(x_0,y_0)$ and $\psi(x_0,y_0)$ respectively, can be represented in various ways, including:
\begin{itemize}
    \item A color representation in which the brightness of each pixel is proportional to $\epsilon(x_0,y_0)$ and the color is a function of  $\psi(x_0,y_0)$ (modulo $\pi$ radians);
    \item A grey-scale representation where the brightness of each pixel is proportional to $\epsilon(x_0,y_0)$;
    \item A color representation in which the color of the pixel is a linear combination of three different hues (e.g.~red, green and blue) for the three components of the diffusion tensor.  Other choices of color space may also be employed.
\end{itemize}

\section{Experiments}

To validate the proposed theoretical approach, experimental x-ray speckle tracking data were collected at the European Synchrotron Radiation Facility (ESRF) in Grenoble, France. The experimental setup corresponds to  Fig.~\ref{fig:ExperimentalSchematic}.
Two distinct experiments were performed. 

The first analysis reuses the data collected and processed in our previous article \cite{MIST1}. These data consist of images of a red currant sample collected at ESRF beamline BM05. The sample was located approximately 55~m downstream of the source, in the beam path of the x-ray photons produced by synchrotron radiation from a 0.85 T dipole bending magnet. The x-ray beam energy was narrowed to a bandwidth of $\Delta E/E \approx 10^{-4}$ at the energy of $E = 17$~keV, using a double Bragg crystal Si(111) monochromator located 27 m from the x-ray source. The speckle generator consisted of a piece of sandpaper with grit size P800 that was fixed on piezo translation motors located 0.5~m upstream of the sample. A FReLoN (Fast Read-Out Low-Noise \cite{labiche1996frelon,douissard2012}) e2V camera, coupled to an optic imaging a thin scintillator, was used to image the sample from a distance $\Delta = 1$~m downstream of it. The effective pixel size of this imaging optical system was 5.8~$\mu$m, with a signal to noise of greater than $500$.

The second experiment employed a similar type of set-up, but on another beamline of the ESRF, which is dedicated to biomedical imaging (ID17).  Here the imaged sample was a mouse knee, with the x-ray photons being produced using a 3 T wiggler. The continuous spectrum of the x-ray source was filtered to a 52 keV narrow energy band by a double bent Silicon crystal monochromator in a Laue-Laue configuration. The speckle-generating diffusive membrane was composed of Cu powder (mean grain size 36 $\mu m$).  This membrane was placed approximately 140 m downstream of the x-ray source, with the sample being placed 1 m downstream of the membrane. The x-ray detector intercepted the beam at a distance $\Delta = 11 m$ beyond the sample. This imaging detector consisted of a SCMOS (PCO 5.5, Germany) camera coupled to an optic imaging a LuAG scintillator, with the full system providing an effective pixel size of approximately 6.31 $\mu m$. While the FreLoN detector was designed for having a higher signal to noise ratio, this second detector design was driven by a higher velocity of read-out and correspondingly lower radiation-dose deposition. Its use eventually resulted in noisier images that were obtained using a fraction of the exposure time that was used for the first sample.  Regarding the mouse sample, the images were obtained several months after the sacrifice of the animal in accordance with Directive 2010/63/EU, with the experiments having been performed in an agreed animal facility (C3851610006) evaluated and authorized by an Ethical Committee for Animal Welfare (APAFIS \#13792-201802261434542 v3).

For both experiments, the set of reference-speckle images $I_{\textrm R}$ was collected in the absence of the sample and by moving the diffusive membrane (either sandpaper or the custom-built membrane composed of Cu powder) to defined positions of the speckle-generator motors. The sample images $I_{\textrm S}$ were acquired while replacing the sandpaper at precisely the same locations with an accuracy on the order of 0.1~$\mu$m. The sets of images were then processed by running a Python3 code on a simple desktop machine. This code is available under a MIT license on a GitHub repository located at the url https://github.com/DoctorEmmetBrown/popcorn. 

\section{Results}

\begin{figure*}
    \centering
    \includegraphics[width=1\textwidth]{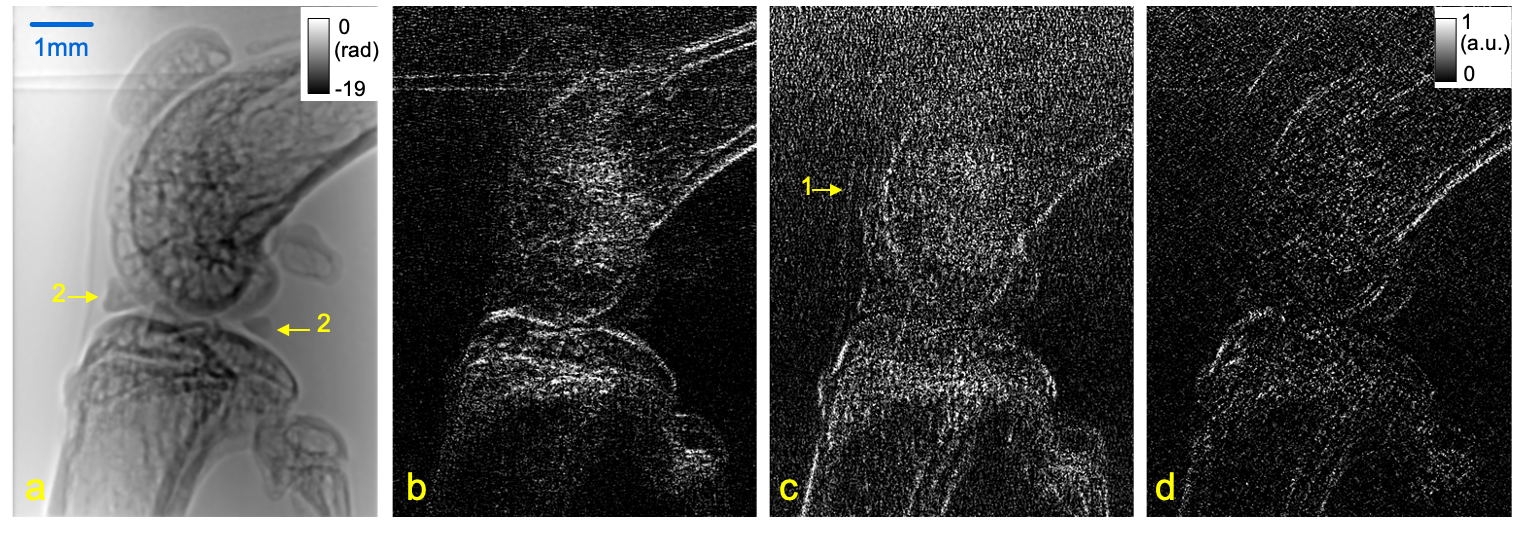}
    \caption{Results for directional dark-field implicit x-ray speckle tracking of a mouse leg from ten pairs of acquisitions $(I_{R},  I_{S})$. (a) Retrieved thickness from MIST converted into a phase map; darkfield tensor components (b) $D^{(yy)}_{\textrm{eff}}$, (c) $D^{(xx)}_{\textrm{eff}}$ and (d) $D^{(xy)}_{\textrm{eff}}$. }
    \label{fig:MiceLeg}
\end{figure*}

Figure~\ref{fig:MiceLeg} presents the results obtained on the mouse knee, in lateral view. We used the first approach (pure phase object) with a $\delta$ value of bone at 52 keV which is equal to
$1.52 \times 10^{-7}$.  We report in this figure the recovered phase map (Fig.~\ref{fig:MiceLeg}(a)) as well as the  three components of the darkfield tensor, i.e~$D^{(yy)}_{\textrm{eff}}$ (Fig.~\ref{fig:MiceLeg}(b)), $D^{(xx)}_{\textrm{eff}}$ (Fig.~\ref{fig:MiceLeg}(c)) and $D^{(xy)}_{\textrm{eff}}$ (Fig.~\ref{fig:MiceLeg}(d)). For better statistics and less noisy results, the displayed darkfield images were computed with more than the four pairs of acquisitions $(I_S, I_R)$ required to solve the system. In this case, ten pairs of speckle images were used, generating due to the noise for each pixel an over-determined system of equations that was solved in the least-squares sense by QR factorization for better numerical accuracy. 
The $\delta$ parameter used for this sample was the value given by http://ts-imaging.science.unimelb.edu.au/Services/Simple/ where data are calculated from the NIST X-Ray Form Factor, Attenuation, and Scattering Tables.

In this figure the different bones can be clearly identified in the phase and the dark-field images, even though the images have a noisy appearance on account of the utilized detector. The three dark-field images (see Figs.~\ref{fig:MiceLeg}(b,c,d,)) show  different signals from each other and are very different compared to the phase image. The edges of the bones seem to create a strong dark-field signal and very few other tissues seem to create a dark-field signal, with the exception of the femoral tendon (indicated by the arrow labeled `1') that is visible in Fig.~\ref{fig:MiceLeg}(c). Interestingly the menisci (indicated by the arrows labeled `2') do not seem to create strong darkfield signals even though they are clearly visible in the phase image. This is probably due to their composition, as menisci are mainly composed of calcified solid cartilage \cite{Broche2021} while the cortical bone is composed of hydroxyapatite crystals arranged to form a porous microstructure. This might be of great interest for osteoarthritis studies as the thickening of calcified cartilage appears to be one of the first signs of this disease. Future studies will be designed to study this phenomenon on a wider range of samples. 

\begin{figure*}
    \centering
    \includegraphics[width=1\textwidth]{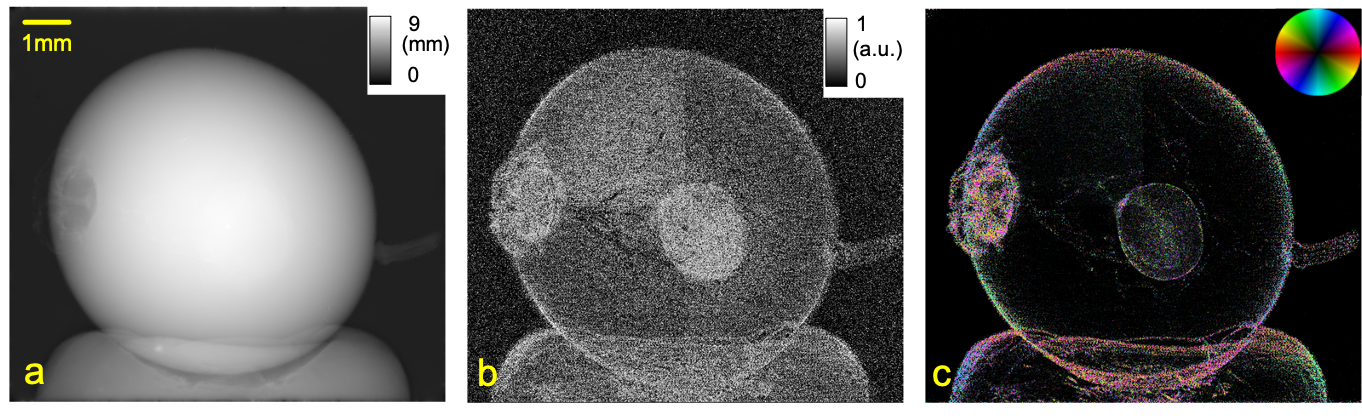}
    \caption{Results for directional dark-field implicit x-ray speckle tracking of a red currant berry from seven pairs of acquisitions $(I_{R},  I_{S})$. (a) Thickness map; (b) eccentricity $\epsilon(x,y)$ of the elliptical SAXS beam; (c) angular orientation $\psi(x,y)$ of the elliptical SAXS beam. The orientations are encoded using a HSV color system with Hue=angle, fixed saturation and brightness value representing the normalized area of the local SAXS ellipse.}
    \label{fig:RedCurrant}
\end{figure*}

Figure \ref{fig:RedCurrant} presents the results of the experiment which used a red currant berry as a sample. We used the second approach (monomorphous material) with $\gamma=1146$ (value for water at 17 keV), using seven pairs of acquisitions. The respective panels show the thickness  map (Fig.~\ref{fig:RedCurrant}(a)), the directional dark-field eccentricity $\epsilon(x,y)$ (Fig.~\ref{fig:RedCurrant}(b)) and the orientation $\psi(x,y)$ of the SAXS-ellipse semi-major axis (Fig.~\ref{fig:RedCurrant}(c)). We again see that the three presented images display complementary information regarding the sample.  In particular, the dark-field quantities plotted in Figs.~\ref{fig:RedCurrant}(b,c) reveal information that is not evident in Fig.~\ref{fig:RedCurrant}(a).  The eccentricity plot in Fig.~\ref{fig:RedCurrant}(b) highlights the oval-shaped seed near the center of the berry, with the interior of this feature being noticeably brighter than the surrounding background.  We also observe higher diffuse-scatter eccentricity at the edges of the sample, which is to be expected on account of our previous comments regarding diffusive flow that is induced by photon scatter from sample edges \cite{YoungOnTheBoundaryWave,Maggi,Rubinowicz,MiyamotoWolf1,MiyamotoWolf2,Groenendijk2020}.  The angular-orientation plot in Fig.~\ref{fig:RedCurrant}(c) exhibits several features that would be expected for a directional dark-field signal \cite{jensen2010a,jensen2010b,yashiro2011}.  The director-field $\psi(x,y)$ clearly traces out the local tangents to the projected edges of the sample, as well as the edges of the supporting mount and the edges of the embedded seed near the center of the sample.  Several fine filaments within the sample also become visible in this director-field plot.  It is also interesting to observe the textured mixture of many angles in the oval-shaped feature to the left of the sample, which is suggestive of an ensemble of unresolved fibrous microstructure with randomly-varying orientations.  Finally, the thickness map of the red currant is quantitative. Indeed, the width of the berry measured from its width on the image and the pixel size is 8.65~mm while the maximum thickness of the berry calculated at its center is 8.71~mm.

\section{Discussion and future work}

\subsection{Discussion}

Our Fokker--Planck speckle-tracking model, for the combined coherent-flow and diffuse-flow channels of x rays traversing a non-crystalline sample, enables us to extract information pertaining to both channels.  This, in turn, relates directly to different aspects of the distribution of the complex refractive index within the sample.  The coherent channel corresponds to a coarse-grained form of the complex refractive index, with the coarse-graining being induced by the finite size of the detector pixels as well as the lack of perfect spatial coherence in the illuminating photon field.  Conversely, the diffuse channel corresponds to a fine-grained form of the complex refractive index, relating to structures that are not directly spatially resolved by the system, but which nevertheless have a measurable influence that may be extracted from recorded maps of radiant exposure.  In this sense, the method may be aptly described using the language of Kagias {\em et al.}~\cite{Kagias2021}, as simultaneously extracting both real-space and reciprocal-space information regarding the sample.

We have given particular emphasis to the recovery of information associated with the diffuse channel, and in particular to the directional-darkfield aspects of this channel.  The closed-form analytic solutions, developed in solving the inverse problem associated with our model, enable computationally rapid darkfield reconstructions using a relatively small number of images. We formulated two different sample models: either a pure phase object or a single-material object. These two models were tested on two different experimental x-ray beamlines with two different samples. The obtained results already seem to show interesting features on the composition of the bone and calcified cartilage that needs to be confirmed with histology.

Next, we briefly discuss our use of the term `darkfield' in the context of our paper.  This term has a more general usage, relating to any form of imaging in which unscattered photons---or other imaging quanta, such as electrons or neutrons---make no contribution to the output image \cite{gage1920}.  This imaging-system property causes samples, which incompletely cover the field of view, to appear with positive contrast against a dark background field.  Thus, when we speak of the darkfield signal in our paper, we are actually referring to a particular form of darkfield signal, namely that which is associated with the diffusion tensors appearing in the anisotropic Fokker--Planck equation (Eq.~(\ref{eq:1--generalised-form})) for speckle tracking.  For a different form of darkfield signal in an x-ray imaging setting, see e.g.~the technique for dark-field x-ray microscopy that is reported in Simons {\em et al.}~\cite{Simons2015}.

It is also worth pointing out a partial analogy that exists, between our technique based on Eq.~(\ref{eq:1--generalised-form}), and the non-equilibrium dynamics of Brownian motion.  In this analogy, consider a thin sheet of ($2+1$)-dimensional fluid in non-uniform flow, at two closely-spaced times $t_1$ and $t_2 > t_1$.  By assumption, this fluid has a position-dependent temperature $T(x,y,t)$, in addition to being anisotropic at a microscopic level, as would be the case e.g.~in a flowing liquid crystal \cite{Jenkins1978}.  Suppose that small clusters of pollen grains have been randomly positioned over the surface of the flowing anisotropic fluid, at time $t_1$.  As the fluid flows, each pollen cluster will be advected.  The timestep $t_2-t_1$ is sufficiently small, by assumption, that the pollen-grain clusters move by  distances no larger than their diameter, in evolving from $t_1$ to $t_2$.  The local displacement of the center of mass of each pollen cluster will be a direct measure of the local velocity of the fluid.  This pollen-cluster advection is analogous to the `speckle tracking' first term on the right side of Eq.~(\ref{eq:1--generalised-form}), if one replaces the randomly-positioned pollen clusters with the randomly-positioned illuminating x-ray speckles, and considers the flow to be induced by placing a transparent sample in the x-ray beam, rather than letting the pollen-laden anisotropic fluid evolve from $t_1$ to $t_2$.  In this analogy, $\Delta \propto (t_2-t_1)$ and $D\propto T$, with the latter fluctuation-dissipation proportionality arising from the Sutherland-Einstein-Smoluchowski relation. In addition to being advected, the pollen clusters will diffuse during the small time interval from $t_1$ to $t_2$, on account of Brownian motion associated with the position-dependent temperature of the fluid.  The temperature distribution, together with the microscopically-anisotropic nature of the fluid film, imply locally-elliptical diffusion of the pollen-grain clusters over small timesteps $t_2-t_1$.  This is analogous to the second and subsequent `anisotropic speckle diffusion' terms on the right side of Eq.~(\ref{eq:1--generalised-form}).   

\subsection{Future work}

Here we list several possible avenues for future work.

\begin{enumerate}

    \item While the formalism and experiment of the present paper was developed in the specific context of x-ray radiation, its Fokker--Planck approach is more broadly applicable to directional dark-field speckle tracking using other parts of the electromagnetic spectrum such as already under study with visible light \cite{Lu2019,wang2019}. Imaging techniques based on alternative particles providing different imaging contrasts like electrons or neutrons \cite{valsecchi2020} could benefit as well from these developments to probe smaller or bulkier samples since one can readily imagine the possibility of developing randomly structured membranes with a broad range of characteristics. 

    \item Following Refs.~\cite{MorganPaganin2019,PaganinMorgan2019}, the symmetric rank-two diffusion tensor in Eq.~(\ref{eq:DiffusionTensor}) may be considered as the first member in an infinite hierarchy of progressively higher-rank tensors, which can be used to extend the speckle-tracking Fokker--Planck equation (Eq.~(\ref{eq:1--generalised-form})) into a Kramers--Moyal form \cite{Risken1989}. Higher-order diffusion-tensor fields relate to progressively higher-order moments of the position-dependent SAXS fans emanating from each point on the exit surface of the sample.  Reference~\cite{PaganinMorgan2019} gives explicit expressions for these higher-order diffusion tensors.  See, also, the earlier papers of Modregger {\em et al.}~\cite{modregger2017,modregger2018}, regarding the role of higher-order SAXS-fan moments in directional dark-field imaging.  

    \item Explicit expressions have been derived for the local-SAXS contribution to the diffusion tensor in Eq.~(\ref{eq:DiffusionTensor}), together with its higher-order Kramers--Moyal generalizations \cite{PaganinLabrietBrunBerujon2018}.  It would be interesting to obtain corresponding  edge-scatter-induced diffusion tensors. Recall, in this context, papers which show DDF arising from sample edges, with the semi-major axis of the diffuse-scatter ellipse being tangential to sample edges \cite{jensen2010b}. This edge-induced DDF signal may be formulated in at least three different ways: (i) Keller's concept of diffracted rays in the geometric theory of diffraction \cite{Keller}; (ii) the Young--Maggi--Rubinowicz boundary-diffraction wave \cite{YoungOnTheBoundaryWave,Maggi,Rubinowicz,MiyamotoWolf1,MiyamotoWolf2}; (iii) critical points of the second kind, resulting from sharp sample edges in asymptotic approximations to diffraction integrals \cite{MandelWolf}.  It is also worth noting, in the context of edge-induced diffuse scatter, that the parabolic equation of paraxial optics is a complex diffusion equation with purely imaginary diffusion coefficient.  
    Another approach to better isolate and quantify the edge dark-field effect would be to utilize the approach in Groenendijk {\em et al.}~\cite{Groenendijk2020} to remove the propagation-based phase contrast edge fringes from the sample image before applying the algorithm described here. Further computation work could look at whether the local stretching of a speckle due to a strong phase gradient can appear as a dark-field signal (see e.g. discussions in Morgan \& Paganin \cite{MorganPaganin2019}).   

    \item Our two-dimensional directional-dark-field reconstructions could be extended to three-dimensional reconstructions, i.e.~tensor dark-field tomography, in an analogous manner to what has already been achieved using periodic-grating methods \cite{Gullberg1999,Malecki2014,bayer2014reconstruction,schaff2015,liebi2015nanostructure,Wieczorek2016}. 

    \item Statistical dynamical diffraction theory  \cite{Kato1,Kato2,PavlovPunegov1998,PavlovPunegov2000,Nesterets2000} would form an interesting perspective from which one might extend the results of the present paper. 
    
    \item The temporal-coherence requirements on x-ray speckle tracking are lax \cite{zdora2015simulations}.  Hence our results might be extended to paraxial polychromatic radiation from sufficiently spatially coherent sources, although the model would likely need in that case to correct for some artifacts observed for instance in Ref.~\cite{vittoria2017retrieving}.
    
    \item As previously mentioned, XSVT and UMPA minimize suitable functionals, with MIST instead solving a particular partial differential equation.  A link between all three speckle-tracking approaches might be explored by recalling the Lagrangian formulation of classical field theory \cite{MandlShawBook}.  Here, minimization of an action-integral functional leads directly to an associated partial-differential equation, namely the Euler--Lagrange equation.  Similarly, minimization of the XSVT and UMPA functionals might lead to a partial differential equation for directional-dark-field x-ray speckle tracking.      
    \end{enumerate}

\section{Conclusion}

We have developed an anisotropic Fokker--Planck equation to perform implicit x-ray speckle tracking.  The method is able to recover the directional dark-field signal associated with spatially unresolved microstructure in a non-crystalline sample.  The method employs illumination of the sample with several spatially-random masks.  The corresponding directional dark-field signals are extracted from the measured radiant-exposure maps, using simple closed-form expressions obtained by solving the inverse problem set up by the Fokker--Planck forward model.  Our theory has been successfully applied to two different experimental data sets, obtained using hard x rays.  We conjecture that these ideas may also be applied to other forms of radiation and matter wave field, such as visible-light photons, electrons and neutrons.      

\section*{Acknowledgments}

We acknowledge the European Synchrotron Radiation Facility for provision of synchrotron radiation facilities. We acknowledge useful discussions with Samantha Alloo, Mario Beltran, Gerard Besson and Thomas Leatham.

\bibliography{MIST2}

\end{document}